# Intelligent Product: Mobile Agent Architecture Integrating the End of Life Cycle (EOL) For minimizing the lunch phase PLM

Abdelhak Boulaalam

Computer science department, Sidi Mohamed Ben AbdEllah University FSDM Fez, LIIAN laboratory
Fez, Morocco
abdelhak.boulaalam@usmba.ac.ma

El Habib Nfaoui

Computer science department, Sidi Mohamed Ben AbdEllah University FSDM Fez, LIIAN laboratory
Fez, Morocco
elhabib.nfaoui@usmba.ac.ma

Omar El Beqqali

Computer science department, Sidi Mohamed Ben AbdEllah University FSDM Fez, LIIAN laboratory
Fez, Morocco
omar.elbeqqali@usmba.ac.ma

*Abstract*— **To improve the increasingly demands products that are customized, all business activities performed along the product life cycle must be coordinated and efficiently managed along the extended enterprise. For this, enterprise had wanted to retain control over the whole product lifecycle especially when the product is in use/recycling (End Of Life phase). Although there have been many previous research works about product lifecycle management in the beginning of life (BOL) and middle of life (MOL) phases, few addressed the end of life (EOL) phase, in particular, when the product is at the customers. In this paper, based on product embedded device identification (PEID) and mobile agent technologies, and with the advent of the development of the ''intelligent products'', we will try to improve innovation: (a) by minimize the lunch phase, (b) and the involvement of the customer in product lifecycle.**

## I. INTRODUCTION

In the actual globally changing business environment, enterprises are in search of new ways of providing additional value to customers and gain a competitive over their competitors. Enterprises are focusing on global management of the whole product lifecycles because nowadays worldwide economic conditions demand they make process changes to stay competitive. These change characteristics are reflected in figure 1.

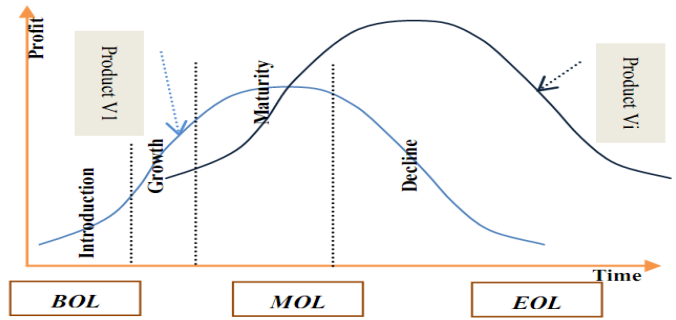

Fig. 1.   Combination biological/industrial lifecycle

The combination between biological and industrial product lifecycle is shown in figure 1. This figure is adapted from [2] and [3]. Past initiatives intended only at product cost, quality, or time-to-market are no longer sufficient to achieve market advantage. The focus today is on innovation [4] [5] [6] [7] [8]. In all these research, to improve innovation throughout companies, the focus is on the intelligent product paradigm that it aims to close the product lifecycle between the end of life and the beginning of life.

There are many projects found in the literature that aims to respond the following points: close the lifecycle, define the intelligent product, improve innovation, and involvement of the customers. Since there many research interests to address the information gap through attempts to gather, analyze and share product related data across extended enterprise. Growth in this area of research has accelerated especially by the maturity and the sharp reduction in the cost curve technologies such as product embedded device identification (PEID). The potency of importance in this field is also demonstrated by research projects such as PROMISE [9], BRIDGE [10] and commercial activities [11] [12] aimed at using AI (Automatic Identification) for collecting product related data and developing systems to manage and use data gathered throughout lifecycle.

In this research we combine with two technologies, multi agents system especially the mobile agent and the PEIDs, to provide more interactivity between all PLM actors and to add the new level of intelligence such as decision making and integrating the end of life phase. For this, we focus on the tree main ideas:

- Creativity because it is the key of innovation;

- Agility because productivity is also the speed (time to market), adaptation of organizations to the quickly change of the environment;

- Cooperation because the gains lie in the way of combining different activities together, even beyond the walls of each company.

After this introduction, Section 2 exposes the review of related studies, the PLM and the closed-loop PLM, and the intelligent product like that the technologies behind this concept. Section 3 describes the architectural model for integrating the end of life phase based on intelligent product. Section 4 gives the discussion and conclusion.

## II. REVIEWS OF RELATED STUDIES

In this section we give an overview of related works and the technologies behind our proposed architecture. First we start with the PLM and closed-loop PLM paradigms, next we give some research based intelligent product, then we discuss the MAS and PEID technologies applied in PLM domain, finally we introduce our proposition based on the previous concepts.

### A. PLM and cosed-loop PLM

Product Lifecycle Management (PLM) is a concept with multiple interpretations. The must common is about management the data generated in the whole product lifecycle. This concept appeared in the late 1990s moving outside the engineering aspects of a product, and providing a shared platform for creation, organization, and dissemination of product-related knowledge across the extended enterprise.

CIMdata [13] defines PLM in four important points:

- A strategic business approach that support the collaborative creation, management, dissemination, and use of product definition information;

- Supporting the extended enterprise;

- Spanning from concept to end of life of a product or plant;

- Integrating people, processes, business systems, and information;

In others terms, CIMdata develops the concept of PLM as knowledge management by gather, storage, processing and usage of product associated knowledge during the life of a product. there many research in the rationale PLM strategies for more and more ability to make better end-of-life decisions, and improve detection of deficiencies or failures in the design or manufacturing process [14] [15], they identified the need to develop a comprehensive platform capable of closing the product information lops from manufacture to its eventual end. Other important definition of the PLM given by Saaksvuori [17] that consider PLM system a collaborative backbone allowing people throughout extended enterprises to work together more effectively.

Reference [3] characterize the life of a product into three distinct phases as shown in Figure 2: beginning, middle, and end:

- (BOL) refers to the time of activities that take place before an asset begins its useable life,

- (MOL) refers to the time of activities,

- (EOL) refers to the time of activities involving recycling and disposal.

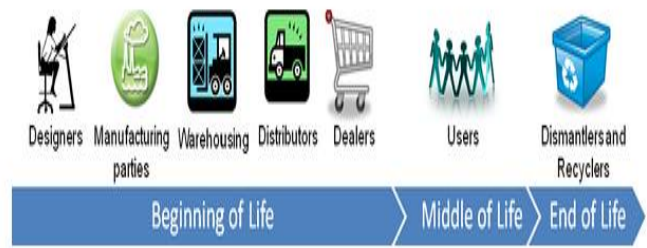

Fig. 2. Actors involved in the various stages of a product's life cycle[3]

Closed-Loop PLM [4] attempts to extend PLM also to the usage, refurbishing, disposal and other lifecycle phases that product instances go through. The focus is in the end of lifecycle. This proposed strategy is based on the emerging technologies, such as RFID, small size sensors and sensor networks or, more generally, product embedded information devices (PEID), a new generation of products called intelligent products are appear. The closed-loop PLM concept shown in figure 2 illustrates the technologies to be explored and developed [16].

### B. Intelligent product

In the literature, the Intelligent Products have many facets, this section is mainly focused on the concept behind Intelligent Products, the technical foundations, and the achievable goals for the Intelligent Product concept will be presented as a starting point for developing the architecture that assure the closed-loop product lifecycle.

Intelligent Products paradigms in [5] have the capabilities to communicate between themselves and with other actors. For this, Intelligent Products link the Auto-ID technology to the agent paradigm and Artificial Intelligence.

McFarlane et al. [18], defines five fundamentals propieties for Intelligent Product:

(1) Possesses a unique identification.
(2) Is capable of communicating effectively with its environment.
(3) Can retain or store data about itself.
(4) Deploys a language to display its features, production requirements, etc.
(5) Is capable of participating in or making decisions relevant to its own destiny.

A product with level 1 product intelligence covers point 1 to 3, a product with level 2 product intelligence covers all points, and is called decision oriented [19]. This intelligent product classification is based on a separation between the actual product (physical product) and its information-based counterpart. For this, in PLM context the use of PEID technology is mandatory.

The fundamental idea behind an Intelligent Product according to [20] is the inside-out control of the supply chain deliverables and of products during their lifecycle. For inside-out control of products, the products should have:

(1) Unique identification code.
(2) Links to information sources about the product across organizational borders.
(3) Even pro-actively

In all this definitions the important is focused on intelligent aspect, for this the classification is given in [18]. This classification on intelligent products is divided on two categories, the first level of Intelligence of Intelligent Products (Information handling, Problem notification, Decision making); this three levels respond to the why question. The second classification aims to respond on the HOW question, in other terms the location of intelligence (Intelligence through network, Intelligence at object, intelligent item, intelligent container).

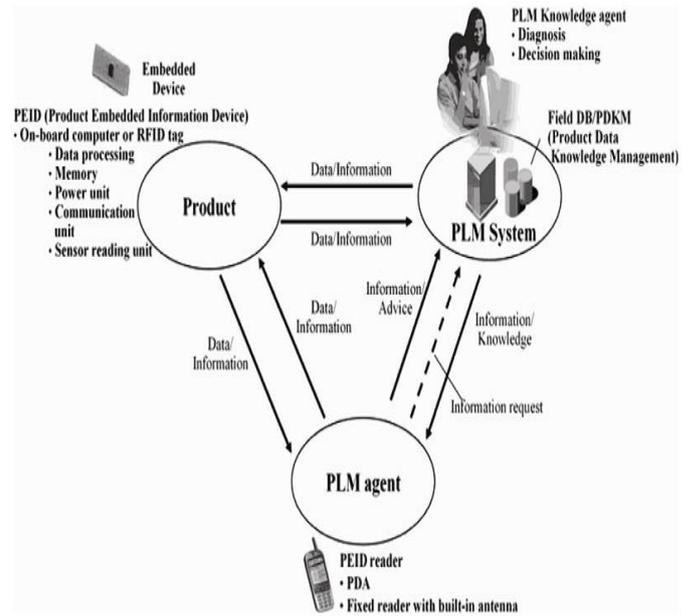

Fig. 3. The closed-loop Product Lifecycle Management concept

Although various definitions of intelligent products have been proposed, we introduce a new definition of the notion of Intelligent Product based on mobile agent and inspired by what happens in nature with us as human beings and the way we develop intelligence and knowledge. The intelligent product concept illustrated in Figure 3. In this model, each product is equipped with a PEID such as RFID tag. this is a novel introduced definition of Intelligent Products, which adds the new level of intelligence by using the mobile agent paradigm.

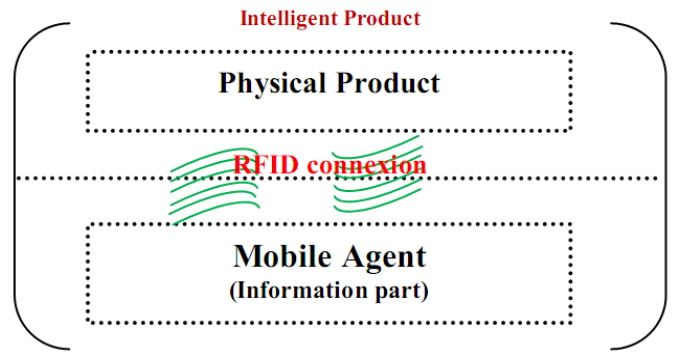

Fig. 4. Intelligent Product

### C. Technologies enabling Intelligent Product

The concept of intelligent product is formulated [21]. In this model, each product is equipped with an Intelligent Data Unit. An Intelligent Data Unit is a hardware device that consists of sensors, a controller, memory, and data communication interface (Figure 5).

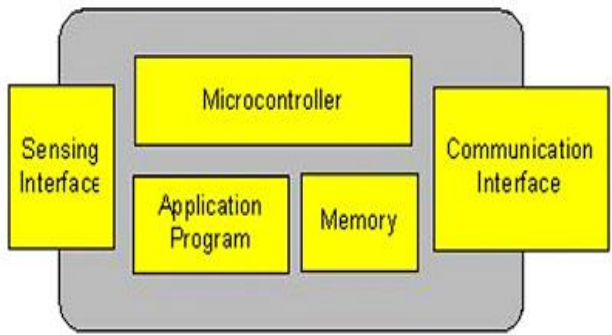

Fig. 5.   Intelligent Data Unit

Auto-ID technologies have been used in the ELIMA [22] and PROMISE projects. In these projects, the aim is to use auto-ID technology to gather dynamic data about product in automatic manner.

The idea of Intelligent Products is to seamlessly connect the products (physical world) with their information counterpart in information systems e.g. through a product agent as proposed in [23].

*D. Multi-agent Systems*

Various projects applied the multi-agent system paradigm to solve different problems in product lifecycle management (simulation, intelligent product, extended enterprise, etc.). Agent technology has already been considered as an important approach for developing industrial distributed systems ([24] [25]).

An information system called the Dialog platform [26] that uses ID@URI was initially developed for tracking products through a supply chain. A product identification and information linking concept labeled ID@URI was proposed at Helsinki University of Technology [27]. DIALOG system is an open source solution developed at the Helsinki University of Technology. DIALOG is a natural implementation platform for the PROMISE [5]. The ID@URI concept and the related DIALOG agent make it possible to query and update product information about tangible things over the internet throughout the product lifecycle. As shown in Figure 6, The DIALOG node contains a simple and configurable mapping mechanism that defines what messages go to which agent(s) and what sender to use for which messages. The proprieties such as protocol, message passing mechanism, security, etc., are implemented by Agents. For this we can say that DIALOG is a ''generic'' software.

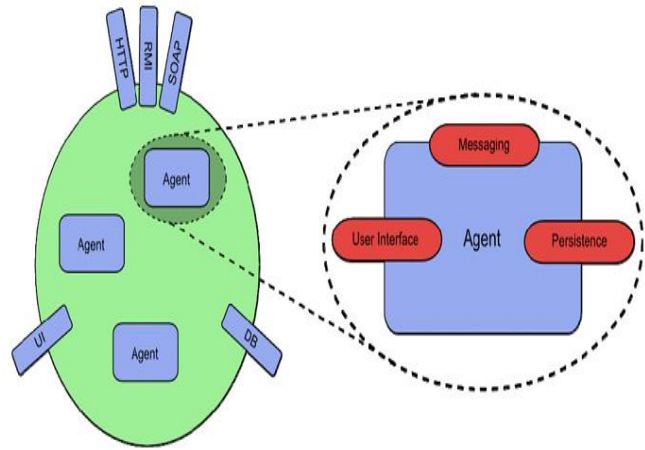

Fig. 6.   DIALOG node and agent

Few of projects founded applied the mobile agent, these entire projects use the stationary agent, and even if this paradigm is used it's not in the product data management domain (e.g. e-Health, e-Commerce etc.). For the e-Health, based on Aglets Software Development Kit (ASDK), a mobile multi-agent based, distributed information platform (MADIP) for wide-area e-health monitoring are proposed to support the intensive and distributed nature of wide-area (e.g., national or metropolitan) monitoring environment [28]. The same project is developed by using the JADE (Java Agent DEvelopment Framework) platform [29].

III.   OUR APPROACH

In the PLM context with the more and more development of technologies, the companies must control his know-how by integrating the product end of life phase. As shown before (Figure 2), the product lifecycle is composed of three phase: begging of life, middle of life, and end of life. We focus on the end of life phase in this paper. According to [3] end of life phase refers to the time of activities involving recycling and disposal of the asset after it has finished its use phase of the lifecycle. Furthermore, in [4] the EOL phase characterized by diverse scenarios (reuse of the product with refurbishing, reuse of components with disassembly and refurbishing, material reclamation without disassembly, material reclamation with disassembly and, finally, disposal with or without incineration).

In our paper we extend the boundaries of the EOL phase, we add the phase of use in this phase. The aim of this extension is the involvement of final customers. Now by using our vision of Intelligent Product, the proposed architecture can improve innovation by optimizing the lunch phase and introduce the new product generation before the end of the ex-product version.

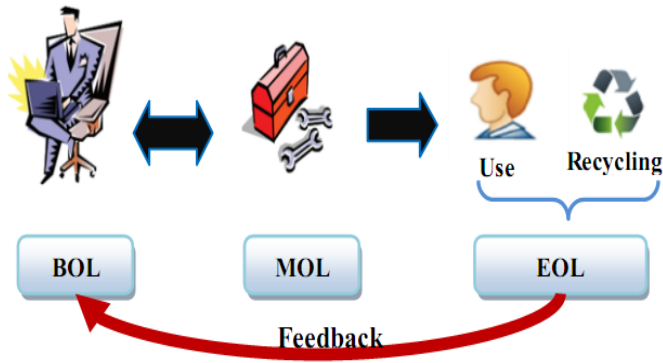

Fig. 7.   Architecture overview

The mobile agents used in our architecture can join and leave the environment without intervention by a system administrator. The AgentProduct and the physical product have the same ID, and with the advent of PEID the AgentProduct is installed on the repair garage, recycling enterprise, and embedded on physical product, and servers.    The overall proposed architecture is depicted in Figure 8.

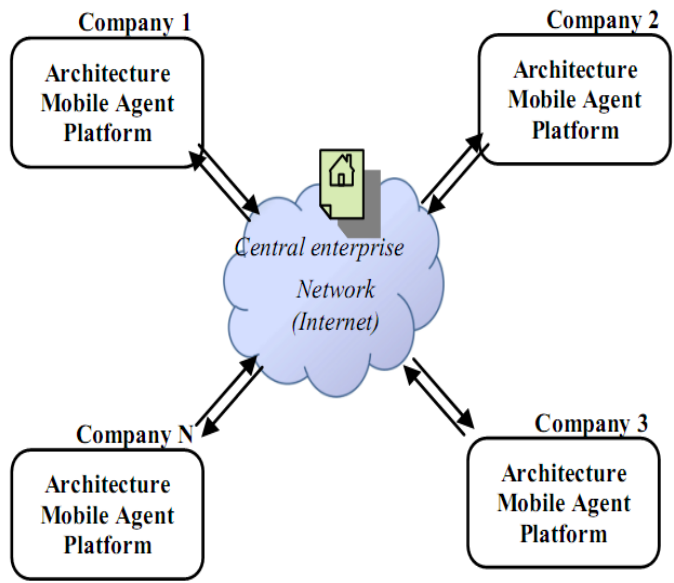

Fig. 8.   The overall framework of proposed architecture

A major challenge in the extended enterprise is how to access the information that is not stored only locally or at only one server but is distributed in two or more places. This is also the key technological challenge in implementing the proposed architecture. By using the intelligent product, the information collected on the product end of life phase can help the designer/manufacturer to improve the design and manufacturing of the products versions, and can be readily personalized to meet current and future customer demands.

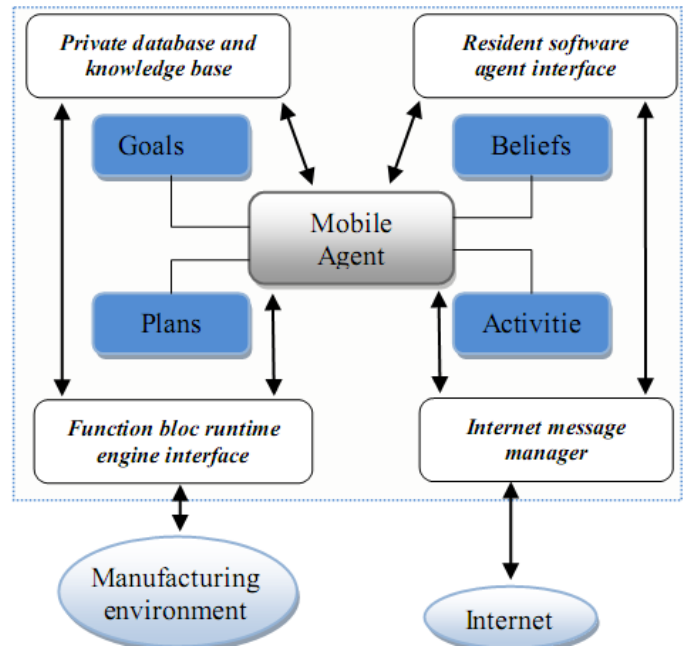

Fig. 9.  Mobile Agent functionalities

To satisfy these requirements, a proposed architecture based on PEIDs and mobile multi-agent architecture that is developed on JADE allows MAs to work on behalf of customer, to collect distributed users' knowledge data, and to spontaneously inform the use gaps situations to associated PLM actors. The components of Mobile Agent model used in our research are illustrated in figure 9 and the class diagram is shown in figure 10.

The complete architecture is shown in Figure 11. This architecture is based on Agent's software that can be mobile in the case of the extended enterprise. Five primary architectural features discriminate the proposed architecture: (1) AgentProduct, (2) AgentService, (3) AgentCustomer, (4) AgentImpact and (5) AgentKnowledge. Figure 10 shows the mobile agent generic implementation level.

In the literatures, knowledge has various dimensions ([30], ([31], ([32], ([33]). In our proposed architecture we aim to present a global view of product lifecycle by integrating the end of life. For this, the knowledge management in this phase is important. We describes below the scenario for creating the knowledge based on the End of Life phase. We categorize the knowledge (1) by the sources (self or collective), (2) the mode (tacit or explicit). Table 1 summarizes some activities and types for knowledge created within the innovation process (non-exhaustive list).

Knowledge and innovation are two closely interconnected and mutually dependent concepts. Generally, the cumulative information/knowledge of a enterprise leads to its innovations. Furthermore, the explicit and tacit knowledge's (in end of life phase) is inseparable and mutually constituted:

(a) the first one (the explicit information/knowledge) is returned by the customer by an interactive participation; for this, we suppose that the product should offer some services such as the detailed operations/uses for new product etc. The customer could be informed that some components are need to be replaced otherwise the product will have many failures. The direct involvement of the customer provides them access to the suitable information, and provides useful information to the manufacturer. (b) The second category is the tacit information/knowledge. Based on the RFID technologies, the detailed information about the product such as the use of product, the environment effects information, and the product failure details can be automatically collected. After this, the AgentProduct based on AgentCustomer, AgentImpact and AgentKnowledge feed the local knowledge repository.

TABLE I.  ACTIVITIES AND TYPES FOR KNOWLEDGE CREATED WITHIN THE INNOVATION PROCESS

| Activity | Knowledge type |
|---|---|
| User insight | Tacit and explicit |
| Market invesigation | Explicit |
| Idea & Concept generation | Tacit and explicit |
| Product requirements | Explicit |
| Engennering & Design | Explicit |
| Merketing & Lunch | Tacit and explicit |
| Sales | Tacit and explicit |
| Custommer | Tacit and explicit |
| Product (Intelligent Product) | Tacit |

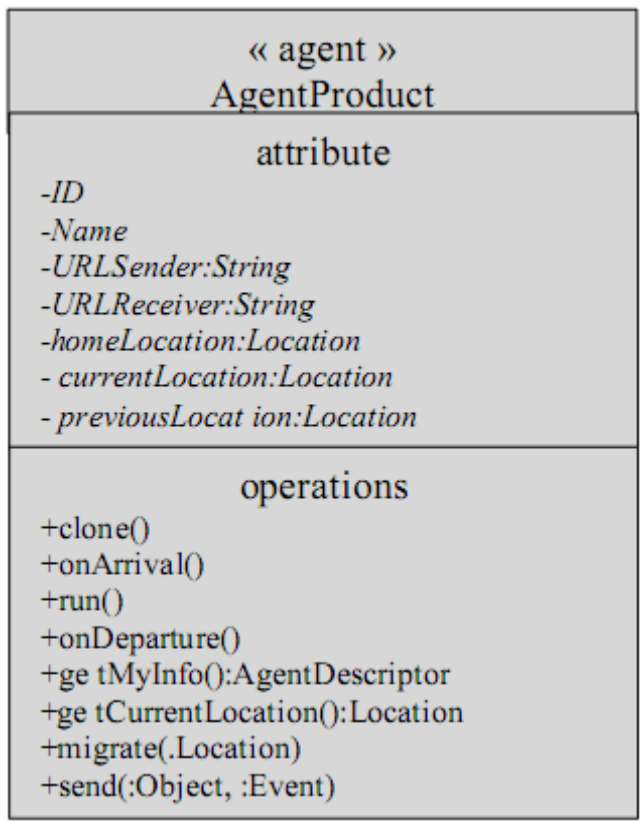

Fig. 10. Example of a class diagram (implementation level)

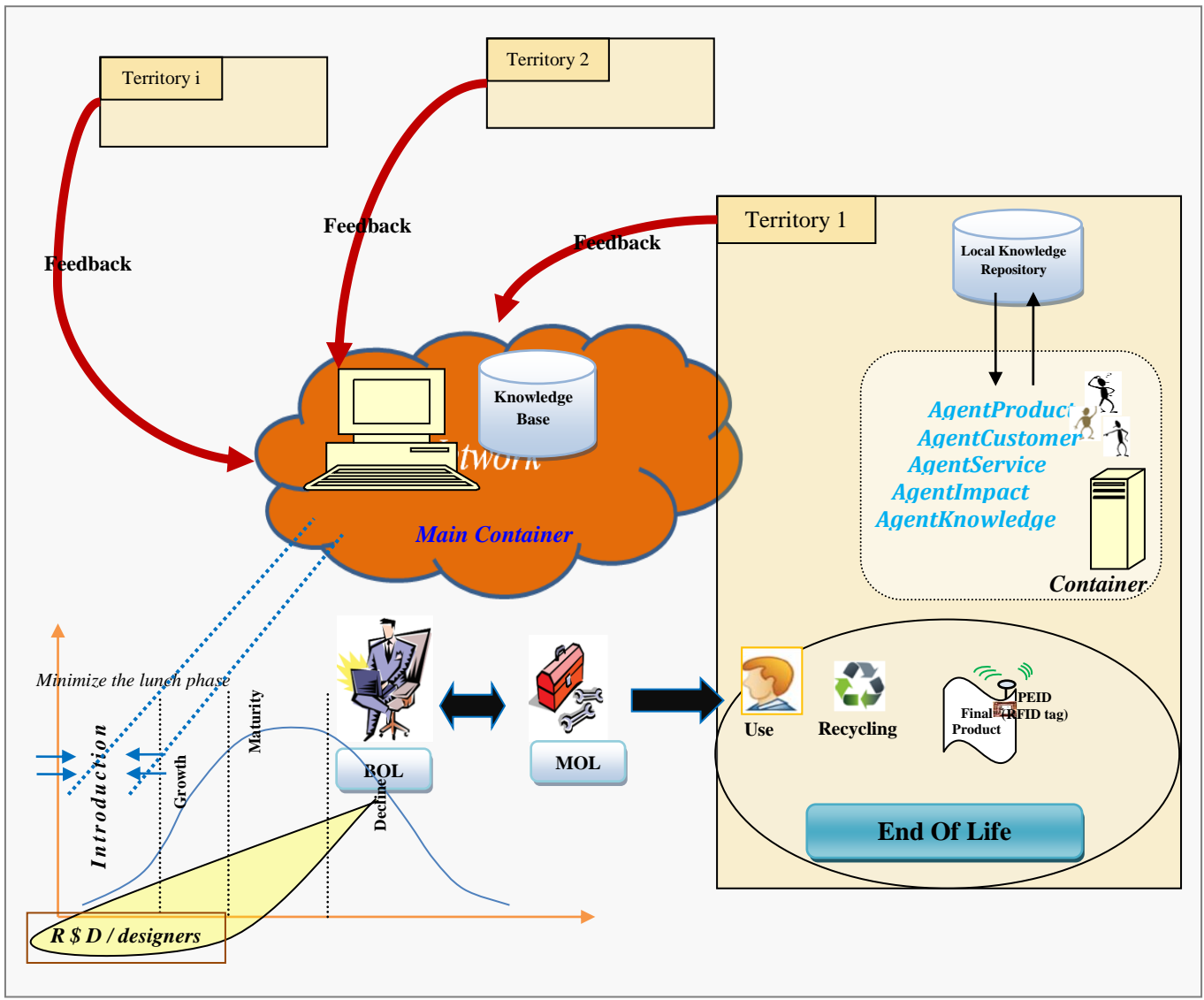

Fig. 11. Our proposed architecture

## IV. CONCLUSIONS

In this study, we provided a new approach focused on the concept of product as a key element (Product in the heart of PLM): the product implicitly embeds the information about itself. For this, it is necessary to add a new dimension to the PLM system in which information flow is horizontally and vertically closed. Starting from the emerging technologies such as product embedded device identification and mobile agent, we proposed an architecture that close the loop product lifecycle management by integrating the end of life phase. For this, we extended the EOL to the use phase. In other terms, this widening aim to implicate the final users in PLM process (involvement of customer). Although this involvement we aim to minimize the lunch phase based on the 'Intelligent Product' feedback: tacit and explicit. Furthermore, in the product use phase, to gather product life cycle data during this phase, the PEID such as RFID has been introduced. Moreover, necessary software components (product information counterpart) and their relations have been addressed: the mobile agent architecture to satisfy the new requirement of extended enterprise. In addition, to streamline several products life cycle operations based on the proposed architecture, the general scenario for creating the knowledge based on the End of Life phase have been introduced.